\documentclass[10pt]{article}
\usepackage{preamble}
\title{A one-dimensional mathematical model for shear-induced droplet formation in co-flowing fluids\footnote[1]{This preprint has not undergone peer review or any post-submission improvements or corrections. The Version of Record of this article is published in Theoretical and Computational Fluid Dynamics, and is available online at https://doi.org/10.1007/s00162-024-00690-5}} %%%%%%%%%%%%
\author[1]{\textbf{Darsh K. Nathawani}\footnote[2]{Email address for correspondence: darshkir@buffalo.edu}}
\affil[1]{Computational and Data-Enabled Science and Engineering, University at Buffalo, Buffalo, NY 14260, USA}
\author[2]{\textbf{Matthew G. Knepley}}
\affil[2]{Department of Computer Science and Engineering, University at Buffalo, Buffalo, NY 14260, USA}

\begin{document}
% \maketitle
\makeatletter
\hfil\parbox[t]{0.7\textwidth}{\centering\LARGE\bfseries\@title}\par
\vspace{1cm}
\hfil\parbox[t]{0.7\textwidth}{\centering\bfseries\@author\\[3ex]\@date}\par
\makeatother
\begin{abstract}
Shear-induced droplet formation is important in many industrial applications, primarily focusing on droplet sizes and pinch-off frequency. We propose a one-dimensional mathematical model that describes the effect of shear forces on the droplet interface evolution. The aim of this paper is to simulate paraffin wax droplets in a co-flowing fluid using the proposed model to estimate the droplet volume rate for different flow velocities. Thus, the study focuses only on the dripping regime. This one-dimensional model has a single parameter that arises from the force balance on the interface.  We use PETSc, an open-source solver toolkit, to implement our model using a mixed finite element discretization. The parameter is defined by cross-validation from previous computational and experimental data. We present the simulation results for liquid paraffin wax under fast-moving airflow with a range of velocities. 
\end{abstract} %%%%%%%%%

\section{{\label{sec:1}Introduction}}

Droplet formation is a naturally occurring process in free-surface flows. However, it can be seen in many scientific and industrial applications, a few examples being ink-jet printing~\cite{Derby2010}, cell bioprinting~\cite{Takagi2019}, spray cooling~\cite{Kim2007}, testing hydrophobic surfaces~\cite{Gao2006}, atomization and droplet entrainment in annular flow~\cite{Inoue_et_al_2021, BernaEscrivaMunozHerranz2015}, and microfluidics devices~\cite{Cramer2004, Teh2008, Dewandre2020microfluidic}.  

A proper understanding of droplet formation started to shape in 1833 when Savart experimented with fluid jets~\cite{Savart1833}. Savart concluded from his experiments that the breakup of a jet always happens regardless of inlet fluid velocity or radius, type of fluid (meaning density and viscosity), or direction of gravitational force. He postulated that it must be due to some intrinsic property of the fluid. But he could not identify it as the surface tension. The role of mean curvature and the surface tension force as a source of instability was explained by Laplace~\cite{Laplace1805} and Young~\cite{Young1805} in 1805. Plateau in 1863 found that the long wavelength perturbations on a fluid jet reduce the surface area, and they become unstable if the wavelength is greater than some critical value. However, he incorrectly concluded that all the wavelengths that reduce the surface energy are unstable~\cite{Plateau1863}. Rayleigh was the first to demonstrate that droplet formation occurs in finite time due to the force of surface tension acting against inertia~\cite{Rayleigh1878}. Hence, in the calculation of the critical wavelength, the fastest growth rate is selected. For the fluid column of radius $h_0$, his linear stability analysis of jet breakup predicted the critical wavelength $\lambda_{cr} = 9.01 h_0$, which agreed well with the experiment by Savart.

After this initial impetus from research in droplet dynamics, the dynamics of a pendant drop received great attention from both mathematicians and scientists. Eggers and Villermaux collected a wonderful record of the advances in understanding the liquid break-up process~\cite{EggersVillermaux2008}. As the drop becomes heavier by continuously adding fluid, gravity overcomes the surface tension and the instabilities start to grow on the interface. The surface tension minimizes the surface energy by decreasing the radius of the fluid column. A spherical droplet starts to form at the end of the fluid column hanging from a very thin neck. Eventually, the radius becomes zero, which is the ``pinch-off'' point, and a droplet separates from the initial fluid column. Immediately after the primary pinch-off, the ``unbalanced'' surface tension creates an impulse on the interface of the neck. The capillary waves due to this recoil perturb the interface before the tip can collapse back to the top of the fluid cone. This leads to a secondary breakup creating smaller droplets, which are called ``satellite drops'', a phenomenon also observed in liquid bridges and decaying jets. 

In 1993, Jens Eggers established a scaling solution of the axisymmetric fluid neck that appears in a droplet breakup process~\cite{Eggers1993}. He explained the universality of the singularity region and the self-similarity of the solution. He also derived the solution of the Navier-Stokes equations before and after the singularity, showing that the solution is characterized by a universal scaling exponent, thus proving the self-similarity~\cite{Eggers1995}. This universal scaling behavior was also explored in singularities of the PDE, especially in interface flows, to explain that the local behavior of the equations is governed by scales that are independent of the initial and boundary conditions~\cite{Constantin1993, Bertozzi1994, Goldstein1993}. Eggers later explained that linear stability analysis provides a reasonable prediction for droplet sizes, although it fails to explain surface evolution once there is sufficient interface deformation~\cite{ChaudharyRedekopp1980a}. This is bound to happen near the singularity due to dominant nonlinear behavior. Moreover, even the higher-order analysis is not able to explain the shape of the drop near singularity~\cite{ChaudharyRedekopp1980a, ChaudharyMaxworthy1980b}.

The research in droplet dynamics is mature~\cite{Eggers1993, Eggers1995, Eggers1997, EggersVillermaux2008} and the mathematics is well-validated by the experiments~\cite{AmbravaneswaranWilkesBasaran2002, ZhangBasaran1995}. For instance, the one-dimensional model using the finite difference method by Eggers and Dupont~\cite{EggersDupont1994} for gravity-induced droplets was sufficiently accurate to describe the primary pinch-off process. The recent investigation of the same model using the finite element method with a self-consistent algorithm was validated with the experiments~\cite{NathawaniKnepleyDropletGravity}. 

Almost all applications of droplet generation require control of certain quantities of interest for the separated droplet, like volume, frequency of droplet generation, surface area, average moving velocity, etc. The emulsion process~\cite{Dewandre2020microfluidic} and atomization process~\cite{BernaEscrivaMunozHerranz2015} come to light when listing the applications of 
shear-force-induced droplet formation. The emulsion process, where the dispersed phase fluid is injected into the continuous phase fluid to produce monodispersed droplets, has been a very interesting matter from both experiments and modeling perspectives. Umbanhowar et al.~\cite {Umbanhowar2000} described an experimental technique to generate a monodisperse emulsion with $\leq 3\%$ polydispersity. Cramer et al.~\cite{Cramer2004} explained the effects of co-flowing ambient fluid on the droplet formation mechanism. Further, they explained that the continuous phase flow velocity, fluid viscosity, and surface tension are the primary factors for droplet formation to happen via the dripping or jetting process. Furthermore, Garstecki et al.~\cite{Garstecki2005} analyzed the droplet breakup in co-flowing immiscible fluids and explained the importance of the geometric confinement on the break-up process. 

Wilkes et al.~\cite{Wilkes1999} performed computational simulations of the evolution of a pendent droplet in ambient quiescent air using two-dimensional equations. They used interface tracking with a finite element approach with an evolving mesh. Their simulation results matched with very high accuracy; capturing length evolution and microthread in a viscous drop. They were able to explain the overturning behavior of the viscous drop before the primary pinch-off. Taassob et al.~\cite{Taassob2017} used the Volume-of-Fluid (VOF) approach with the finite volume method to simulate droplet break-up in co-flowing fluids. They reported that the continuous phase flow rate has a significant impact on the primary droplet sizes. Sauret and Shum used the level-set method with finite elements to investigate jet break-up in microfluidic co-flowing devices. They presented that introducing initial perturbations to the dispersed phase velocity can lead to faster jet break-up and more control over the droplet sizes. Dewandre et al.~\cite{Dewandre2020microfluidic} proposed a non-embedded co-flow-focusing nozzle design that can generate droplets with a wide range of radii without needing any surfactant or coating. They also provided the numerical simulations using Arbitrary Lagrangian-Eulerian (ALE) method. They reported that the droplet size can be controlled entirely by the continuous phase flow velocity for a given dispersed phase fluid in a given geometric set-up.

The previous work on numerical investigation of shear-induced droplets was done using a two-dimensional domain to account for both fluids. In contrast, the droplet formation in the quiescent background was studied with the one-dimensional approach and the assumption of rotational symmetry. The observation of rotational symmetry can also be true in the co-flowing fluids. However, the attempt to devise a one-dimensional mathematical model to provide the shear force effects of the continuous phase on the droplet formation process is yet to be made. In this paper, we endeavor to develop a solution approach to derive a one-dimensional model that may be used to simulate droplet formation in co-flowing fluids. The purpose of this paper is to propose a mathematical model to include the shear force effect of the continuous phase flow on the interface such that the model equations are simplified. We explore the mathematical understanding of the continuous phase flow near the interface and the contribution to the interface evolution. 

This mathematical model is derived in section~\ref{sec:2} with the explanation of the numerical procedure to solve the model equations. The model introduces a parameter that is governed by the shear layer thickness, which can be difficult to foresee. Therefore, we perform cross-validation to define this parameter and validate this one-parameter model with the experiments, which is clarified in section~\ref{sec:3}. We also explore droplet evolution for different continuous phase velocities in the dripping regime.  

%%%%%%%%
\section{{\label{sec:2}Modeling}}

\begin{figure}[!b]
    \centering
    \includegraphics[width=0.5\textwidth]{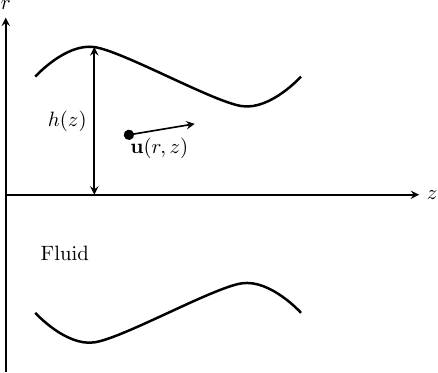}
    \caption{Generic flow geometry describing the interface $h(z)$ and rotational symmetry around the $z$ axis.}
  \label{fig:flow_geometry}
\end{figure}

A one-dimensional approach for droplet formation in a quiescent background was proposed by Jens Eggers and Todd Dupont in~\cite{EggersDupont1994}. Using their approach in a co-flowing fluid configuration, we start with the Navier-Stokes equations in cylindrical coordinates. A generic sketch of the flow geometry is shown in Fig.~(\ref{fig:flow_geometry}). Rotational symmetry is assumed; therefore, the interface can be defined in the $r-z$ plane. The continuity and momentum equations in cylindrical coordinates are given below, with $u$ the velocity of the droplet fluid, $p$ its pressure, and $\nu$ the kinematic viscosity.

\begin{align}
\frac{\partial u_r}{\partial r} + \frac{\partial u_z}{\partial z} + \frac{u_r}{r} &= 0 \label{eq:continuity}\\[2.5ex]
\frac{\partial u_r}{\partial t} + u_r \frac{\partial u_r}{\partial r} + u_z \frac{\partial u_r}{\partial z} + \frac{1}{\rho} \frac{\partial p}{\partial r} - 
\nu \left( \frac{\partial^2 u_r}{\partial r^2} +  \frac{\partial^2 u_r}{\partial z^2} + \frac{1}{r} \frac{\partial u_r}{\partial r} - \frac{u_r}{r^2} \right) &= 0 \label{eq:momentum_1}\\[2.5ex]
\frac{\partial u_z}{\partial t} + u_r \frac{\partial u_z}{\partial r} + u_z \frac{\partial u_z}{\partial z} + \frac{1}{\rho} \frac{\partial p}{\partial z} - 
\nu \left( \frac{\partial^2 u_z}{\partial r^2} +  \frac{\partial^2 u_z}{\partial z^2} + \frac{1}{r} \frac{\partial u_z}{\partial r} \right) - g &= 0 \label{eq:momentum_2}
\end{align}

The radius $r$ is bounded by $h(z,t)$, which is the advecting boundary of the droplet. Since the interface is moving with the flow velocity, it can be modeled using a front-tracking method. The model equation is given by Eq.~(\ref{eq:h_front}).

\begin{align}
  \frac{\partial h}{\partial t} + u_z \frac{\partial h}{\partial z} = u_r \Big|_{r=h} \label{eq:h_front}
\end{align}

\begin{figure}[!t]
    \centering
    \includegraphics[width=0.4\textwidth]{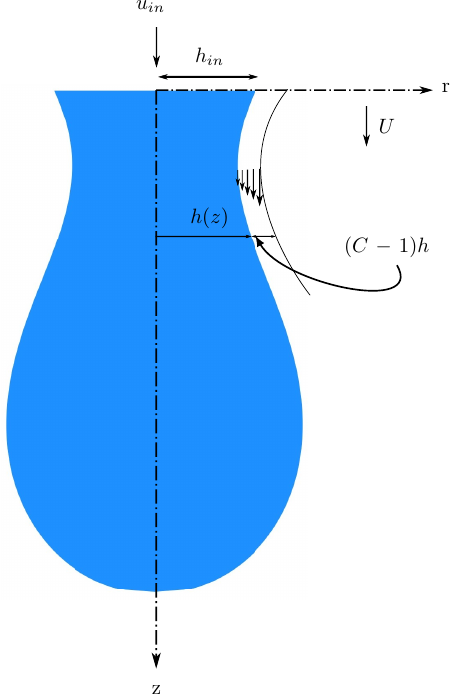}
    \caption{Schematic of a droplet formation in a co-flowing fluid.}
  \label{fig:coflow_schematic}
\end{figure}

Figure (\ref{fig:coflow_schematic}) shows a schematic of a flow configuration where two fluids are flowing in the same direction. In this \textit{co-flowing} configuration, The inner fluid is considered to be the dispersed phase fluid, and a superscript `$d$' is used to define its variables. The outer fluid is called the continuous phase fluid and the variables are defined using a superscript `$c$'. In the case of a co-flow type environment, the velocity profile for the continuous phase flow can be derived by assuming a laminar and incompressible flow. In this section, we first derive the equation for the velocity profile. 
\subsection{Dispersed phase flow}
We define the field variables of the dispersed phase flow using asymptotic expansion. We use the fact that the radius of the fluid column is much smaller than the length. Moreover, the radial contraction is faster than the elongation. Therefore we expand the field variables, namely axial velocity ($u_z^d$) and pressure ($p^d$) using an asymptotic expansion in $r$. Because of the symmetry, we get
\begin{align}
  u_z^d &= u_0^d + u_2^d r^2 + ... \label{eq:asymptotic_uz} \\[2.5ex]
  p^d &= p_0^d + p_2^d r^2 + ... \label{eq:asymptotic_p}
\end{align}
and using $u_z^d$ in continuity equation, we get $u_r^d$.
\begin{align}
    u_r^d &= - \frac{\partial u_0^d}{\partial z} \frac{r}{2} - \frac{\partial u_2^d}{\partial z} \frac{r^3}{4} - ... \label{eq:asymptotic_ur}
\end{align}

We use Eq.~(\ref{eq:asymptotic_uz} - \ref{eq:asymptotic_ur}) to simplify Eq.~(\ref{eq:momentum_2}) for lowest order in $r$.
\begin{align}
  \frac{\partial u_0^d}{\partial t} +  u_0^d \frac{\partial u_0^d}{\partial z} + \frac{1}{\rho} \frac{\partial p_0^d}{\partial z} - \nu^d \left( 4u_2^d + \frac{\partial^2 u_0^d}{\partial z^2} \right) - g = 0 \label{eq:1D_pre_final}
\end{align}

\subsection{Continuous phase flow}
The important aspect of modeling the shearing effects of the continuous phase flow is the description of the velocity profile. We start with the Navier-Stokes equations using cylindrical coordinates given by Eq.~(\ref{eq:continuity}-\ref{eq:momentum_2}) and apply them in the pipe-flow scenario. The radial velocity is assumed to be zero and the axial velocity is assumed to be a function of radius only. Therefore, the momentum equations reduce to
\begin{align}
  \frac{\partial p^c}{\partial r} &= 0 \label{eq:2_21}\\[3ex]
  \frac{1}{r} \frac{\partial}{\partial r} \left( r \frac{\partial u_z^c}{\partial r} \right) &= \frac{1}{\mu^c} \frac{\partial p^c}{\partial z} \label{eq:2_22}
\end{align}
which explains that the pressure does not change along the radial direction. Assuming the constant pressure gradient in the axial direction, we can get the velocity profile as given below.

\begin{align}
  u^c_z = \frac{r^2}{4 \mu^c} \frac{dp}{dz} + C_1 \ln(r) + C_2 \label{eq:u_c_z}
\end{align}

The constants $C_1$ and $C_2$ can be found using the flow conditions of the continuous phase. At the interface, the velocity of the dispersed and continuous phase flow should be equal. The velocity away from the interface should be the same as a Poiseuille flow velocity profile. The distance away from the interface where this condition is satisfied depends on the thickness of the boundary layer in the continuous phase fluid as shown in Fig.~(\ref{fig:coflow_schematic}).

\begin{align}
    r = h \qquad & u^c_z = u^d_z \label{eq:condition_1}\\[1.5ex]
    r = Ch \qquad & u^c_z = \frac{1}{4 \mu^c} \frac{dp}{dz} \left( C^2 h^2  - R^2 \right) \label{eq:condition_2}
\end{align}

Here, the parameter $C$ must be greater than $1$ and $R$ is the radius of the capillary pipe. The parameter $C$ defines the thickness of the shear layer in the continuous phase flow, which is precisely $(C-1)h$. Intuitively, the thickness is some $x\%$ of the interface distance $h$ from the axis of symmetry, where $x$ is $\mathcal{O}(1)$. Therefore, the parameter $C$ must be greater than 1. The thickness of this shear layer defined how much force is experienced by the dispersed phase droplet~\cite{NathawaniPhDThesis}.

Using the above conditions we can get two equations for our two constants. Using the asymptotic expansion for $u^d_z$ from equation (\ref{eq:asymptotic_uz}), we get
\begin{align}
  \frac{h^2}{4\mu^c}\frac{dp}{dz} + C_1 \ln(h) + C_2 &= u^d_0 \label{eq:condition_1_1}\\[3ex]
  C_1 \ln(Ch) + C_2  &= - \frac{R^2}{4\mu^c} \frac{dp}{dz} \label{eq:condition_2_1}
\end{align}
We subtract equation (\ref{eq:condition_1_1}) from equation (\ref{eq:condition_2_1}) to get $C_1$. Then we substitute $C_1$ in equation (\ref{eq:condition_1_1}) to get $C_2$.
\begin{align}
  C_1 &= \frac{\left( h^2 - R^2 \right)}{4 \mu^c \ln(C)} \frac{dp}{dz} - \frac{u^d_0}{\ln(C)} \label{eq:C1}\\[3ex]
  C_2 &= u^d_0 + u^d_0 \frac{\ln(h)}{\ln(C)} - \frac{h^2}{4 \mu^c} \frac{dp}{dz} - \frac{\left( h^2 - R^2 \right) \ln(h)}{4 \mu^c \ln(C)} \frac{dp}{dz} \label{eq:C2}
\end{align}
Substituting these constants in equation (\ref{eq:u_c_z}) yields the equation for the continuous phase velocity.
\begin{align}
  u^c_z = u^d_0 + u^d_0 \frac{\ln(h/r)}{\ln(C)} + \frac{\left(r^2 - h^2 \right)}{4\mu^c}\frac{dp}{dz} - \frac{\left( h^2 - R^2 \right)}{4 \mu^c} \frac{dp}{dz} \frac{\ln(h/r)}{\ln(C)} \label{eq:uc}
\end{align}
The axial velocity profile given in the equation (\ref{eq:uc}) is the proposed velocity profile in the region close to the interface. The radial velocity of the continuous phase is assumed to be the same as the dispersed phase radial velocity at the interface to satisfy the kinematic condition.

\subsection{Force balance}
The force balance can be performed at the interface using the proposed velocity profile. The stresses from the dispersed and continuous phase flow contribute to the interface dynamics in both normal and tangential directions. The total normal forces are balanced by the surface tension force and the total tangential forces are zero.
\begin{align}
  \hat{\mathbf{n}} \sigma^d \hat{\mathbf{n}} + \hat{\mathbf{n}} \sigma^c \hat{\mathbf{n}} &= - \gamma \mathcal{K} \label{eq:balance_n} \\[3ex]
  \hat{\mathbf{n}} \sigma^d \hat{\mathbf{t}} + \hat{\mathbf{n}} \sigma^c \hat{\mathbf{t}} &= 0 \label{eq:balance_t}
\end{align}
where $\sigma^c$ and $\sigma^d$ represent the stress tensor for the continuous phase and dispersed phase respectively. The normal and tangent vectors are defined by
\begin{align}
\mathbf{\hat{t}} &= \frac{1}{\sqrt{1 + \left(\frac{\partial h}{\partial z}\right)^2}} \begin{bmatrix}
\frac{\partial h}{\partial z}\\ 1
\end{bmatrix} \label{eq:t} \\[1.5ex]
\mathbf{\hat{n}} &= \frac{1}{\sqrt{1 + \left(\frac{\partial h}{\partial z}\right)^2}} \begin{bmatrix}
1 \\ -\frac{\partial h}{\partial z}
\end{bmatrix} \label{eq:n}
\end{align}

The surface tension coefficient is given by $\gamma$ and $\mathcal{K}$ is the curvature, which is defined by
\begin{align}
  \mathcal{K} = \left[ \frac{1}{h \left (1 + \frac{\partial h}{\partial z}^2\right )^{1/2}} - \frac{\frac{\partial^2 h }{\partial z^2}}{\left (1 + \frac{\partial h}{\partial z}^2\right )^{3/2}} \right] \label{eq:curvature}
\end{align}

The stresses in the normal and tangential directions are given by
\begin{align}
    \hat{\mathbf{n}} \sigma \hat{\mathbf{n}} &= - p + \frac{2\mu}{\left[1+\left(\frac{\partial h}{\partial z}\right)^2\right]} \left[ \frac{\partial u_r}{\partial r} + \frac{\partial u_z}{\partial z} \left(\frac{\partial h}{\partial z}\right)^2 - \left(\frac{\partial u_z}{\partial r} + \frac{\partial u_r}{\partial z}\right) \frac{\partial h}{\partial z} \right] \label{eq:stress_n} \\[2ex]
    \hat{\mathbf{n}} \sigma \hat{\mathbf{t}} &= \frac{\mu}{\left[1+\left(\frac{\partial h}{\partial z}\right)^2\right]} \left[ 2 \frac{\partial u_r}{\partial r} \frac{\partial h}{\partial z} + \left(\frac{\partial u_z}{\partial r} \frac{\partial u_r}{\partial z}\right)   +  \left(1 - \left(\frac{\partial h}{\partial z}\right)^2\right) - 2\frac{\partial u_z}{\partial z} \frac{\partial h}{\partial z} \right] \label{eq:stress_t}
\end{align}

We use Eqs.~(\ref{eq:stress_n}) and (\ref{eq:stress_t}) in the force balance given by Eqs.~(\ref{eq:balance_n}) and (\ref{eq:balance_t}). The force balance can be simplified using the proposed axial velocity profile given in the equation (\ref{eq:uc}). The radial velocity of the continuous phase is the same as the interface velocity in the vicinity of the interface. Therefore, we can use equation (\ref{eq:asymptotic_ur}) for the radial velocity. The expansion of the force balance provides a set of equations to derive the governing equations. 

In the case of the tangential force, we get $\mathcal{O}(h^{-1})$ and $\mathcal{O}(h)$ terms, which provide us with two equations.
\begin{align}
\frac{u_0 \mu^c}{h \ln(C)} + \frac{R^2}{4 h \ln(C)}\frac{dp^c}{dz} = 0 \qquad \qquad\label{eq:Ft_1}\\[3ex]
4 u_2 = \frac{6}{h}\frac{\partial u_0}{\partial z}\frac{\partial h}{\partial z} \left( 1 + \frac{\mu^c}{\mu^d} \right) + \frac{\partial^2 u_0}{\partial z^2}\left( 1 + \frac{\mu^c}{\mu^d} \right)
 - \frac{1}{\mu^d}\frac{dp^c}{dz} - \frac{1}{2\mu^d \ln(C)}\frac{dp^c}{dz} \label{eq:Ft_2}
\end{align}

Similarly, we collect $\mathcal{O}(1)$ terms from the normal force balance and use the equation (\ref{eq:Ft_1}) for simplification.

\begin{align}
  - p_0 - \mu^d \frac{\partial u_0}{\partial z} &= - p^c + \mu^c \frac{\partial u_0}{\partial z} - \gamma \mathcal{K} + 2 \frac{\partial h}{\partial z} \cancelto{0}{\left( \frac{u_0 \mu^c}{h \ln(C)} + \frac{R^2}{4 h \ln(C)}\frac{dp^c}{dz} \right)} \nonumber \\[3ex]
  \therefore p_0 &= p^c - \mu^d \frac{\partial u_0}{\partial z} \left( 1 + \frac{\mu^c}{\mu^d} \right) + \gamma \mathcal{K}  \label{eq:Fn}
\end{align}

We can use the Eq.~(\ref{eq:Ft_2}) and (\ref{eq:Fn}) to substitute corresponding terms in the equation (\ref{eq:1D_pre_final}). We drop the subscripts and superscripts for dispersed phase velocity, $u_0^d$, for the simplistic appearance of the equations. Also, we add the buoyancy force as an external force to the system, which yields the following momentum equation. 
\begin{align}
  \frac{\partial u}{\partial t} +  u \frac{\partial u}{\partial z} + \frac{\gamma}{\rho} \frac{\partial \mathcal{K}}{\partial z} &- \frac{6 \nu^d}{h} \frac{\partial u}{\partial z} \frac{\partial h}{\partial z} \left( 1 + \frac{\mu^c}{\mu^d} \right) 
  - 3 \nu^d \frac{\partial^2 u}{\partial z^2}\left( 1 + \frac{2}{3}\frac{\mu^c}{\mu^d} \right)  \nonumber \\[2.5ex]
  &+ \frac{2}{\rho^d}\frac{dp^c}{dz} + \frac{1}{2 \rho^d \ln(C)}\frac{dp^c}{dz} - \left( 1 - \frac{\rho^c}{\rho^d} \right) g = 0 \label{eq:co-flow_momentum}
\end{align}
where, $\mathcal{K}$ is the curvature defined by equation (\ref{eq:curvature}).

The interface equation given by Eq.~(\ref{eq:h_front}) can be simplified using the asymptotic expansion of the radial and axial velocities given by Eqs.~(\ref{eq:asymptotic_ur}) and (\ref{eq:asymptotic_uz}). This simplification yields the following equation for interface tracking.
\begin{align}
  \frac{\partial h}{\partial t} + u \frac{\partial h}{\partial z} + \frac{h}{2}\frac{\partial u}{\partial z} = 0 \label{eq:interface}
\end{align}

\subsection{Mixed finite element approach}

The Eqs.~(\ref{eq:co-flow_momentum}) and (\ref{eq:interface}) are solved to simulate the droplet formation of an inner fluid in a two-fluid system with a given outer flow velocity by Eq.~(\ref{eq:uc}). These are solved for $u$ and $h$ using a finite element discretization. However, the highest order derivative is of the third order, which is problematic for our $C^0$ continuous element scheme. The approximation for this term will be discontinuous across element interfaces. Inspired by Ambravaneswaran et al.~\cite{AmbravaneswaranWilkesBasaran2002}, we choose a mixed-element formulation, in which we explicitly discretize the axial derivative of the radius $h$ with a new field variable $s$. The equation for this new variable is given by
\begin{align}
  s - \frac{\partial h}{\partial z} = 0 \label{eq:s}
\end{align}
The curvature term in equation (\ref{eq:curvature}) is now augmented by the new field variable $s$ such that
\begin{align}
  \mathcal{K} = \left[ \frac{1}{h \left (1 + s^2\right )^{1/2}} - \frac{\frac{\partial s}{\partial z}}{\left (1 + s^2\right )^{3/2}} \right] \label{eq:augmented_curvature}
\end{align}

We derive the weak form using this mixed-form variable in the governing equations. Performing the integration by parts, the mixed finite element formulation is given by
\begin{align}
  &\int_{\Omega} q \left[ \frac{\partial u}{\partial t} +  u \frac{\partial u}{\partial z} -\frac{6\nu}{h}\frac{\partial h}{\partial z}\frac{\partial u}{\partial z}\left( 1 + \frac{\mu^c}{\mu^d} \right) + \frac{\gamma}{\rho} \left\{-\frac{s \frac{\partial s}{\partial z}}{h \left (1 + s^2\right )^{3/2}}  - \frac{s}{h^2 \left (1 + s^2\right )^{1/2}}\right\}  \right. \nonumber  \\[1.5ex]
  &\qquad \left.   + \frac{1}{2 \rho^d \ln(C)}\frac{dp^c}{dz} - \left( 1 - \frac{\rho^c}{\rho^d} \right)g \right] d\Omega  + \int_{\Omega} \nabla q \left[3 \nu \left( 1 + \frac{2}{3}\frac{\mu^c}{\mu^d} \right) \frac{\partial u}{\partial z}\right.  \nonumber\\[2ex]
  &\qquad \left. + \frac{\gamma}{\rho} \frac{\frac{\partial s}{\partial z}}{\left (1 + s^2\right )^{3/2}}  \right] d \Omega   - \int_{\Gamma} q \left[3 \nu \left( 1 + \frac{2}{3}\frac{\mu^c}{\mu^d} \right) \frac{\partial u}{\partial z} + \frac{\gamma}{\rho} \frac{\frac{\partial s}{\partial z}}{\left (1 + s^2\right )^{3/2}}  \right] d \Gamma = 0   \label{eq:FE_momentum}  \\[3ex]
  &\int_{\Omega} v \left [ \frac{\partial h}{\partial t} + u \frac{\partial h}{\partial z} + \frac{h}{2} \frac{\partial u}{\partial z} \right ]d \Omega = 0 \label{eq:FE_interface} \\[3ex]
 &\int_{\Omega} w \left [ s - \frac{\partial h}{\partial z} \right ] d \Omega = 0  \label{eq:FE_s}
\end{align}
Here, $q$, $v$, and $w$ are test functions, and the mean curvature is defined by equation (\ref{eq:augmented_curvature}).

The discrete system given by Eqs.~(\ref{eq:FE_momentum} - \ref{eq:FE_s}) is solved using Galerkin finite elements using a third order polynomial approximation for $u$ and $h$ and second order for $s$.

\subsection{Stabilization}
The Galerkin formulation presents a limitation on the stability of the convection-dominated problems. The droplet formation problem is inherently convective in nature. Therefore, in addition to the Galerkin approximation, there is a need to have stabilization in the formulation. The discrete form of the Galerkin formulation for the convection-dominated problems neglects a truncation error that is dissipating~\cite{DoneaHuerta2003}. This neglected dissipation term causes the Galerkin approximation to produce a solution with negative dissipation. To compensate for this negative dissipation, the artificial diffusion term is added in the formulation using Streamline Upwinding (SU) scheme. The convective term in the strong form is
\begin{align}
  \overbrace{\left( u - \frac{6 \nu^d}{h} \frac{\partial h}{\partial z} \right)}^{u_{conv}}\ \frac{\partial u}{\partial z} \label{eq:convective_term}
\end{align}
The artificial diffusion term that corresponds to this convective operator is given by
\begin{align}
  \overbrace{\beta \frac{u_{conv}\ \Delta z }{2}}^{\bar{\nu}}\  \frac{\partial^2 u}{\partial z^2} \label{eq:artificial_diffusion}
\end{align}
where, $u_{conv}$ is convective velocity, $\Delta z$ is the element size, and $\beta$ is the parameter that governs the amplitude of the added diffusion. The value of $\beta$ is chosen between 0 and 1. $\beta = 1$ corresponds to the full upwind differencing scheme and $\beta = 0$ corresponds to zero added diffusion. The use of a fully upwinding scheme is discouraged~\cite{GreshoLee1979, DavidMallinson1976} due to its excessively dissipative results. Therefore, $\beta = 0.5$ is used for the simulations in this study. This stabilization was only enabled for low-viscosity fluids, like paraffin wax, water, etc. High-viscosity fluids like glycerol can be handled without any stabilization.

Another approach for a stable formulation is to use Steamline Upwiding Petrov-Galerkin (SUPG) method~\cite{BrooksHughes1982}. In SUPG formulation, the strong form of the equations is regularized using a similar parameter as $\bar{\nu}$. However, the strong form equation has a third order derivative of $h$ from the curvature term. Using the SUPG stabilization scheme defeats the purpose of using the mixed finite element formulation.
% \vfill
%
\subsection{Initial and boundary conditions}
Initially, the curvature profile is a hemisphere since that minimizes surface energy, which is given by $\sqrt{h_0^2 - z^2}$. The slope ($s$) is the derivative of this equation. The initial velocity is prescribed as zero. \\[3ex]
\noindent\textbf{Initial conditions:} \label{sec:IC}
\begin{align*}
    h &= \sqrt{h_0^2 - z^2} \\[2ex]
    s &= -\frac{z}{\sqrt{h_0^2 - z^2}}\  \text{for}\ (0\leq z < L_0),\qquad s|_{L_0} = - S \\[2ex]
    u &= 0
\end{align*}
where S is a large negative number. In our implementation, we use $-10$. However, the code was tested with larger values and the results were unchanged. 

The inlet radius $h_0$ is fixed depending on the nozzle radius and the inflow velocity $u_0$ is constant. The radius at the tip of the droplet, at length $L(t)$, is zero. The set of Eqs.~(\ref{eq:FE_momentum})-(\ref{eq:FE_s}) are then solved using a continuous Galerkin formulation subject to the following constraints on the boundaries. \\[3ex]
\noindent\textbf{Boundary conditions:} \label{sec:BC}
\begin{center}
\setlength{\tabcolsep}{12pt}
\begin{tabular}{l | l l}
 $z = 0$   & $h = h_0$ & $u = u_0$ \\[2ex]
 $z = L(t)$  & $h = 0$   & $u = \frac{dL}{dt}$
\end{tabular}
\end{center}
\vspace{0.5cm}

The length of the drop L(t) can be calculated as a part of the solution as explained by Ambravaneswaran et al.~\cite{AmbravaneswaranWilkesBasaran2002} by calculating the volume of the drop, which can then be used to calculate the velocity at the tip. However, this results in a dense row in the Jacobian, so we instead produce $L(t)$ by self-consistent iteration which is explained in our previous  work~\cite{NathawaniKnepleyDropletGravity}.

The non-dimensional numbers defined below are used for defining the parameter $C$ and the analysis of the results. A total of five dimensionless quantities are used for the analysis and are given as follows:
\begin{enumerate}
  \item The velocity ratio of continuous phase to dispersed phases, $\frac{u^c}{u^d}$.
  \item The viscosity ratio of continuous phase to dispersed phases, $\frac{\mu^c}{\mu^d}$.
  \item Weber number of the dispersed phase ($We^d$), which is a measure of the importance of the inertial forces over the surface tension forces.
  \begin{align}
    We^d = \frac{\rho^d {u^d}^2 h_0}{\gamma}
    \label{eq:We}
  \end{align}
  \item Capillary number of the continuous phase ($Ca^c$), which is a ratio of the viscous drag force vs surface tension force.
  \begin{align}
    Ca^c = \frac{\mu^c u^c}{\gamma}
    \label{eq:Ca}
  \end{align}
  \item Ohnesorge number of the dispersed phase ($Oh^d$), which relates viscous forces to inertial and surface tension forces.
  \begin{align}
    Oh^d = \frac{\mu^d}{\sqrt{\rho^d \gamma h_0}}
    \label{eq:Oh}
  \end{align}
\end{enumerate}

\section{Analysis and results}\label{sec:3}

\subsection{Validation}

The most important factor for the shear-induced droplet model is the definition of the parameter $C$. Hence, the cross-validation is done to define the parameter $C$ using the findings by Hua et al.~\cite{Hua2007}. Then this one-parameter model is validated using the defined parameter $C$ by comparing the droplet profile with the experimental results by Cramer et al.~\cite{Cram1984}. The quantities we use to define $C$ are the velocity ratio and viscosity ratio of the continuous phase to the dispersed phase, and the surface tension by comparing the effects of these quantities on the droplet radius. Intuitively, the shear layer thickness is some $x\%$ of the interface distance from the axis of symmetry $h$, where $x$ is $\mathcal{O}(1)$. Therefore, the parameter $C$ must be greater than 1. The shear force exerted by the continuous phase fluid on the dispersed phase droplet is embedded in the thickness of this shear layer.

\begin{figure}[!t]
    \centering
    \begin{subfigure}[t]{0.45\linewidth}
        \centering
        \includegraphics[width=\textwidth]{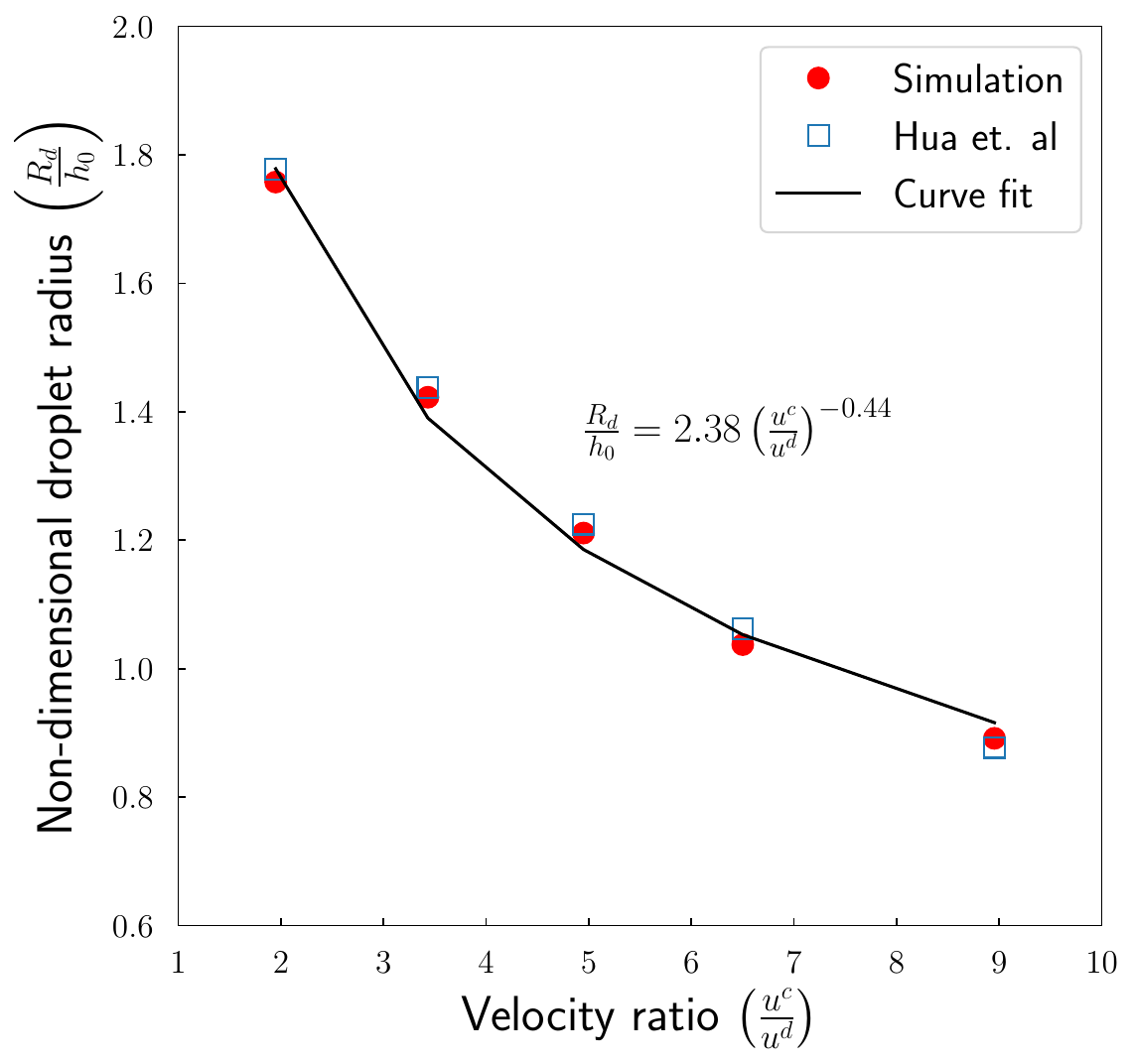}
        \caption{}
        \label{fig:validation_u}
    \end{subfigure}
    \hfill
    \begin{subfigure}[t]{0.45\linewidth}
        \centering
        \includegraphics[width=\textwidth]{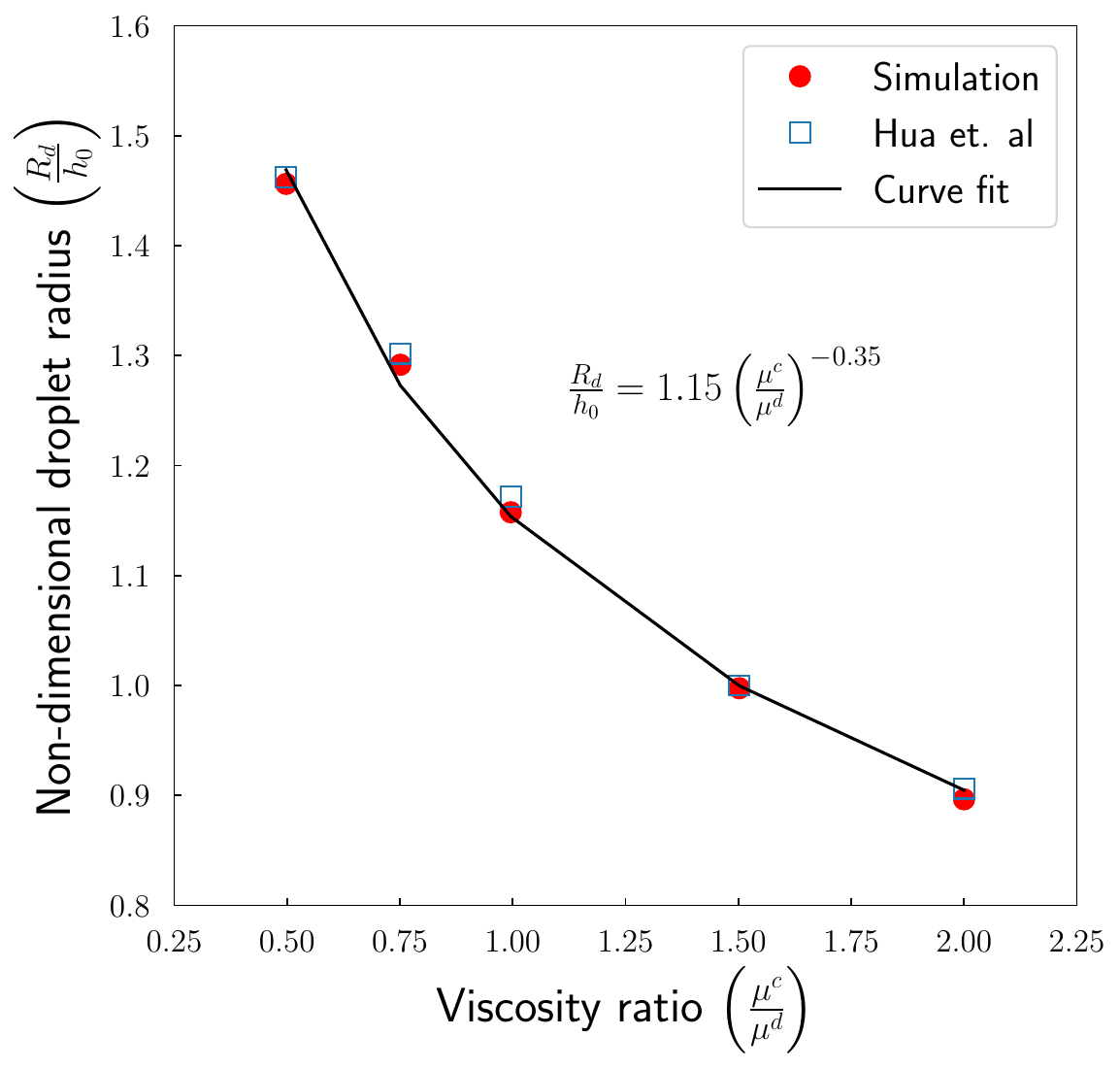}
        \caption{}
        \label{fig:validation_mu}
    \end{subfigure}
    \begin{subfigure}[t]{0.45\linewidth}
        \centering
        \includegraphics[width=\textwidth]{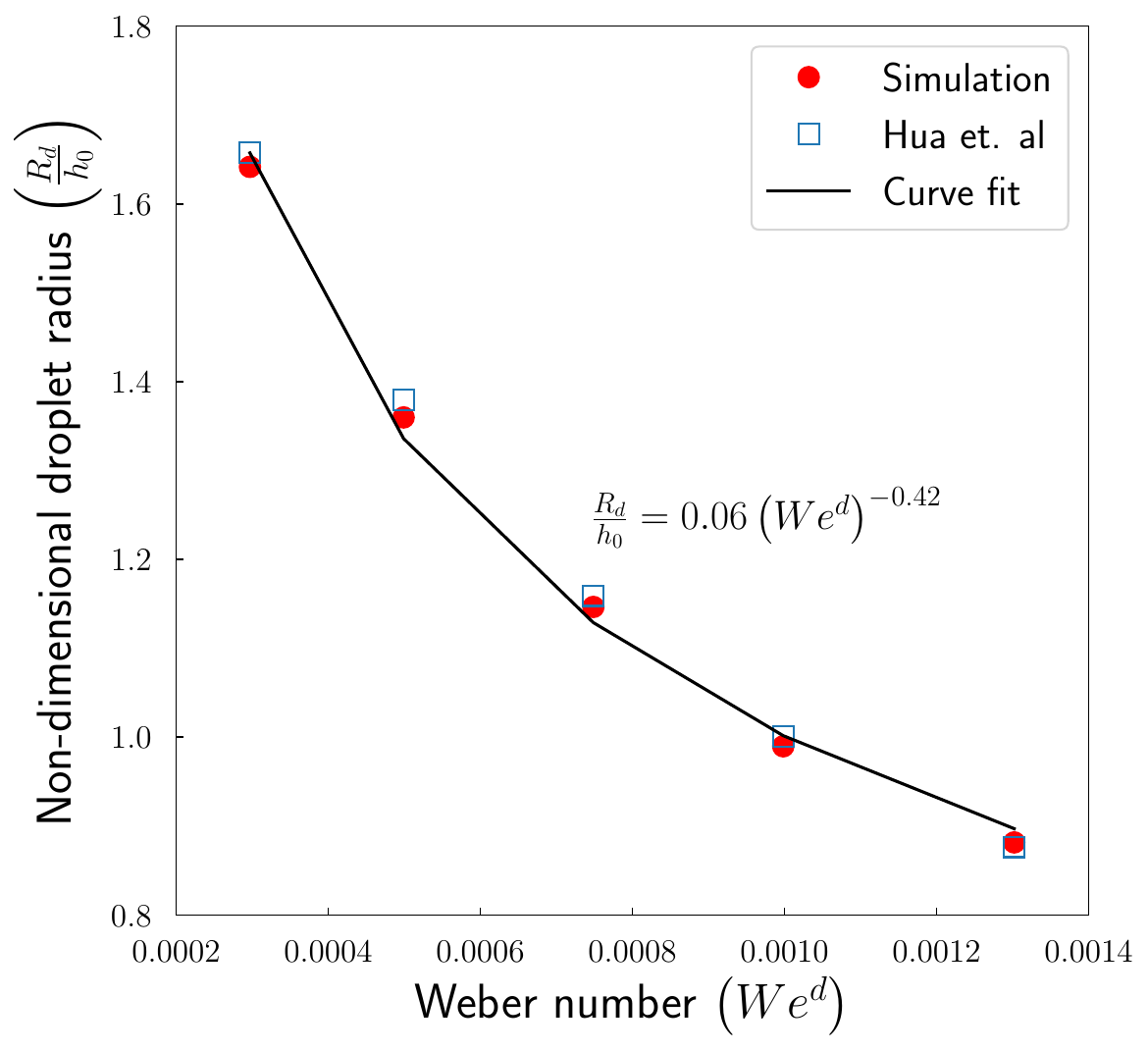}
        \caption{}
        \label{fig:validation_gamma}
    \end{subfigure}
    \caption{Change in non-dimensional radius for (a) different velocity ratios of continuous to dispersed phase, (b) different viscosity ratios of continuous to dispersed phase, and (c) different surface tension forces.}
    \label{fig:enter-label}
\end{figure}

As reported by Hua et al.~\cite{Hua2007}, the droplet radius is affected by the viscosity and velocity ratios of two fluids and the surface tension force at the common interface for given initial conditions. Primarily, this suggests that the balance of viscous, inertial, and surface tension forces between two fluids define the coefficient for the thickness of the interface layer, which is $(C-1)$. Since the parameter $C$ is dimensionless, we can define the dependence of this parameter on dimensionless quantities. The dimensionless quantities used are the Capillary number of the continuous phase flow $Ca^c$, the Ohnesorge number of the dispersed phase flow $Oh^d$, and the viscosity ratio of the continuous and dispersed fluids. The functions to define $C$ are chosen to be monotonic, $tanh()$, and $exp()$,  to have the inclusive lower bound. We tune this parameter using these monotonic functions such that it verifies the dependence on velocity and viscosity ratios given in Fig~(\ref{fig:validation_u}) and (\ref{fig:validation_mu}). We then validate this tuned parameter by comparing the results for different Weber numbers, which is shown in Fig.~(\ref{fig:validation_gamma}). The function is given by
\begin{align}
  C &= 1 + \left[ 0.45\tanh(2.50\ Oh^d - 2.0) + 0.45 \right] \nonumber \\[1.5ex]
  & \left[ 20.0\exp\left( -45 Ca^c\ \left(\frac{\mu^c}{\mu^d}\right)^{-0.6} + 0.045 \right) \right]
  \label{eq:C}
\end{align}
%
% \begin{figure}[!t]
%   \centering
%   \includegraphics[width=0.75\linewidth]{3.pdf}
%   \caption{Change in non-dimensional radius for different velocity ratios of continuous to dispersed phase.}
%   \label{fig:validation_u}
% \end{figure}
% %
% \begin{figure}[!h]
%   \centering
%   \includegraphics[width=0.75\linewidth]{4.pdf}
%   \caption{Change in non-dimensional radius for different viscosity ratios of continuous to dispersed phase.}
%   \label{fig:validation_mu}
% \end{figure}
% %
% \begin{figure}[!h]
%   \centering
%   \includegraphics[width=0.75\linewidth]{5.pdf}
%   \caption{Change in non-dimensional radius for different surface tension forces.}
%   \label{fig:validation_gamma}
% \end{figure}
%

The change in non-dimensional droplet radius for various velocity ratios is shown in Fig.~(\ref{fig:validation_u}). The plot compares the findings of Hua et al. and the simulation results using the Eq.~(\ref{eq:C}). The simulation results provide a very good agreement with the comparison data. The droplet radius for increasing velocity ratio gets smaller. Similarly, Fig.~(\ref{fig:validation_mu}) describes that the non-dimensional droplet radius gets smaller as the viscosity ratio increases.  This suggests that the droplets get smaller as the shear force from the continuous phase fluid gets larger. Also, the impact of change in surface tension force on the droplet radius is reported in Fig.~(\ref{fig:validation_gamma}). The Weber number, as given by Eq.~(\ref{eq:We}), is inversely proportional to the surface tension force. Therefore, the non-dimensional droplet radius decreases for increasing the Weber number or decreasing surface tension force. This suggests that higher surface tension force creates bigger droplets due to the higher regularization of the curvature derivative.

\begin{figure}[!b]
    \centering
    \includegraphics[width=0.4\linewidth]{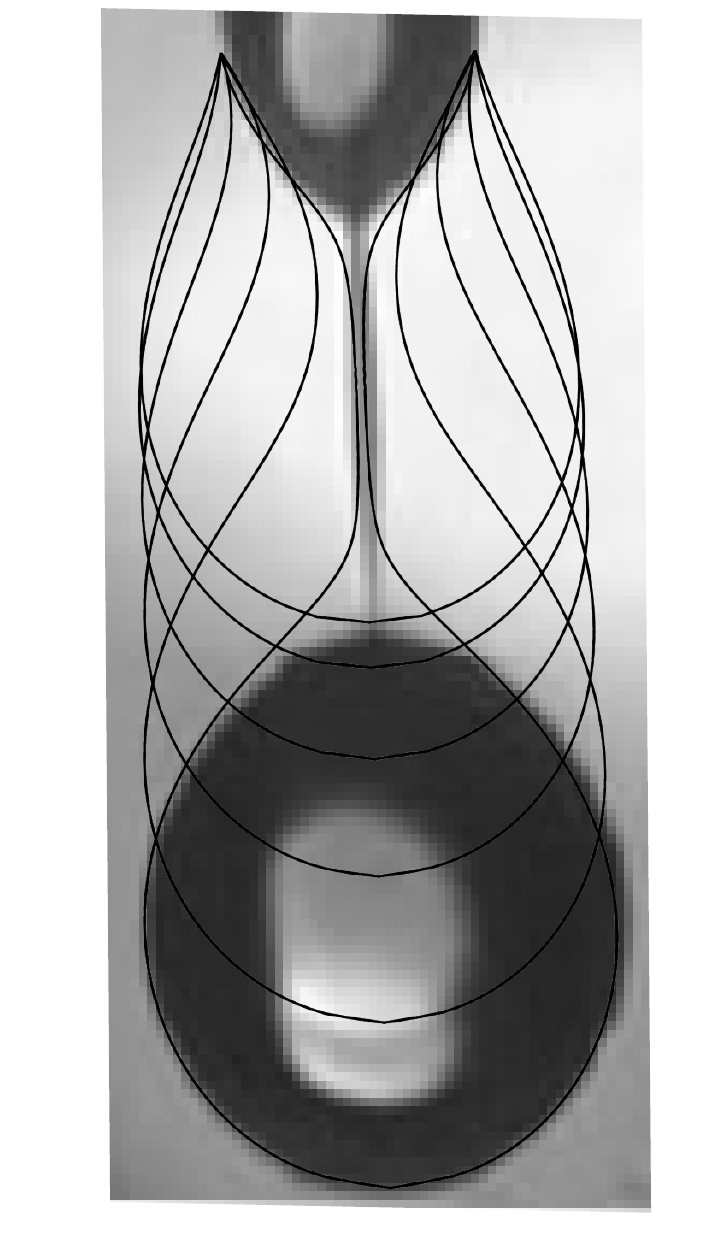}
    \caption{Comparison of the shear-induced droplet experiment done by Cramer et al.~\cite{Cramer2004} and simulation results using the defined parameter $C$ by Eq.~(\ref{eq:C}).}
    \label{fig:Validation_droplet}
\end{figure}

The empirical relation that defines the parameter $C$ by Eq.~(\ref{eq:C}) can be used to perform simulations for different materials and flow conditions. However, comparing the experimental data can be useful in gaining confidence in the defined correlation. Therefore, a simulation is performed using the experimental findings by Cramer et al.~\cite{Cram1984}. The comparison is shown in Fig.~(\ref{fig:Validation_droplet}) for $68\%$ the mixture of $\kappa$-Carrageenan in water as the dispersed phase fluid and sunflower oil as the continuous phase fluid. The continuous phase velocity is $0.075$ m/s. Other fluid properties are used as given in~\cite{Cramer2004}. The droplet profile from the simulation is overlapped with the experimental image. The comparison shows that the one-dimensional model can correctly identify the length evolution of the droplet. However, some discrepancy is seen in the curvature profile in the region of high curvature gradient. The simulated curvature near the top and bottom regions of the neck deviates from the actual curvature. This discrepancy might be because the shear force is assumed to be valid across the entire interface with the prescribed viscosity and velocity profile. However, this difference is only seen for the high viscosity ratio of the continuous and dispersed fluids. The primary droplet profile is still predicted well, with reasonable accuracy near pinch-off.

\subsection{Paraffin Wax droplets}

In this section, the simulations are performed for a liquid paraffin wax droplet in the co-flow environment where the continuous phase fluid is air. The paraffin wax properties are used from the previous work~\cite{NathawaniKnepleyDropletGravity}. The simulations are done using varying continuous phase velocities. The inlet radius ($h_0$) is $1.6$ mm and the inlet flow velocity of the dispersed phase ($u^c$) is $0.01$ m/s. The radius of the outer capillary ($R$) is considered to be $3h_0$ to be consistent with the validation results. The simulation cases are run only for shear force and in the absence of gravity. The continuous phase velocity range is chosen from $7$ m/s to $15$ m/s with an increment of $1$ m/s, keeping the jet's evolution in the dripping regime.

\begin{figure}[!hbt]
  \centering
    \begin{subfigure}[t]{0.45\linewidth}
      \centering
      \includegraphics[width=\textwidth]{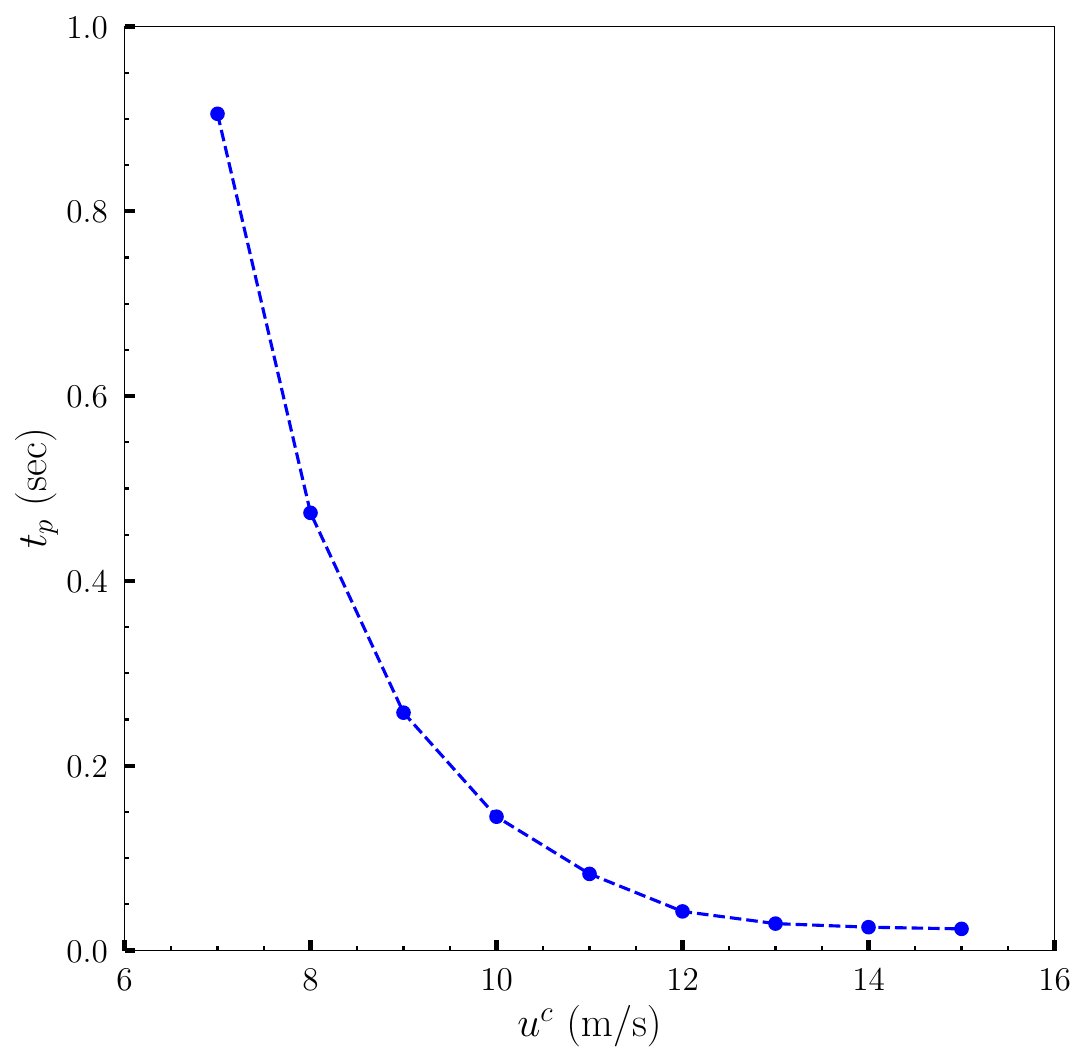}
      \caption{}
      \label{fig:wax_t_pinch_shear}
    \end{subfigure}
    \begin{subfigure}[t]{0.45\linewidth}
      \centering
      \includegraphics[width=\textwidth]{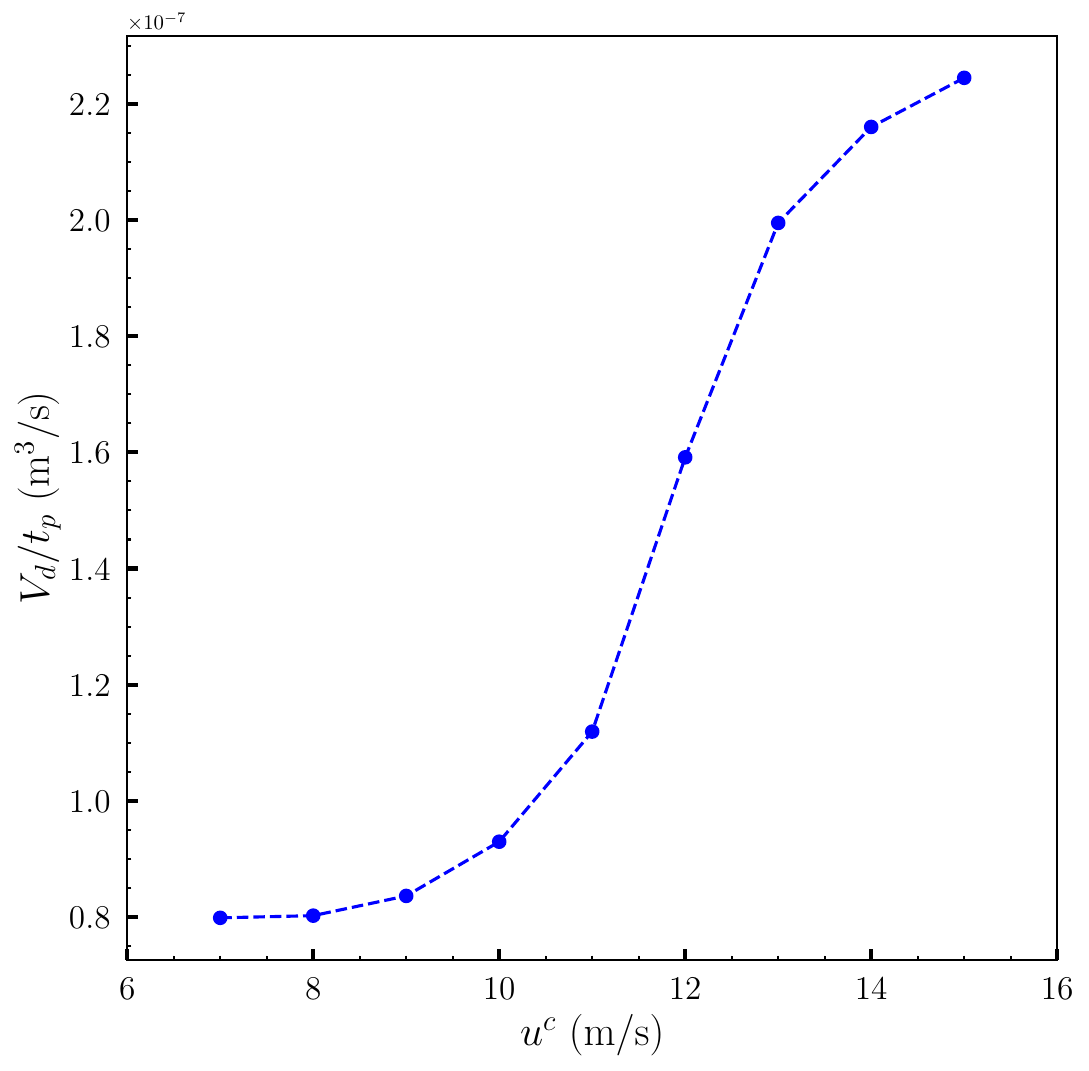}
      \caption{}
      \label{fig:wax_V_t_shear}
    \end{subfigure}
    \caption{(a) Pinch-off time and (b) Pinch-off volume of the primary droplet per unit time for various continuous phase velocities for shear-induced liquid paraffin wax droplet.}
\end{figure}
\begin{figure}[!hbt]
% \ContinuedFloat
  \centering
    \begin{subfigure}[t]{0.45\linewidth}
      \centering
      \includegraphics[width=\textwidth]{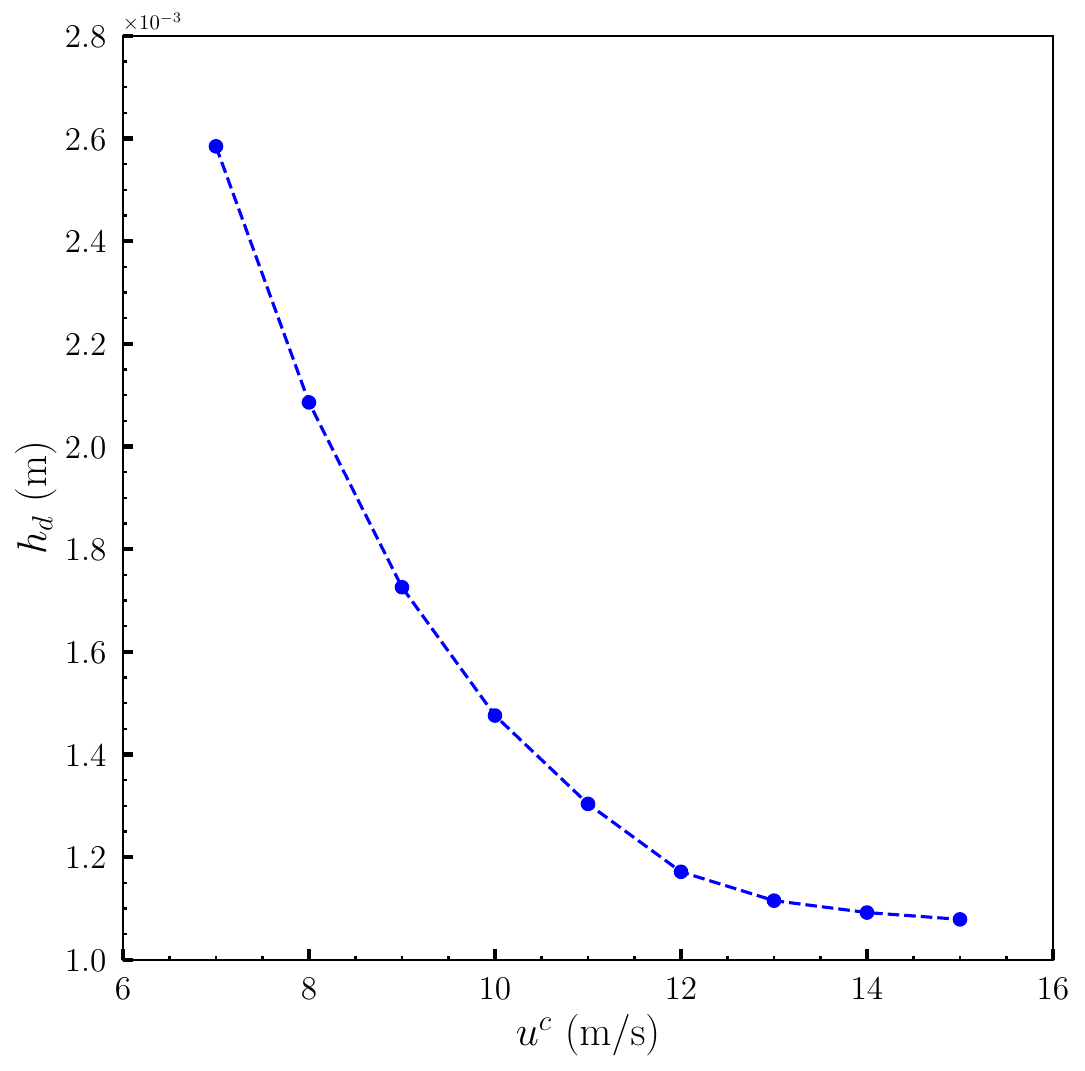}
      \caption{}
      \label{fig:wax_h_drop_shear}
    \end{subfigure}
    \begin{subfigure}[t]{0.45\linewidth}
      \centering
      \includegraphics[width=\textwidth]{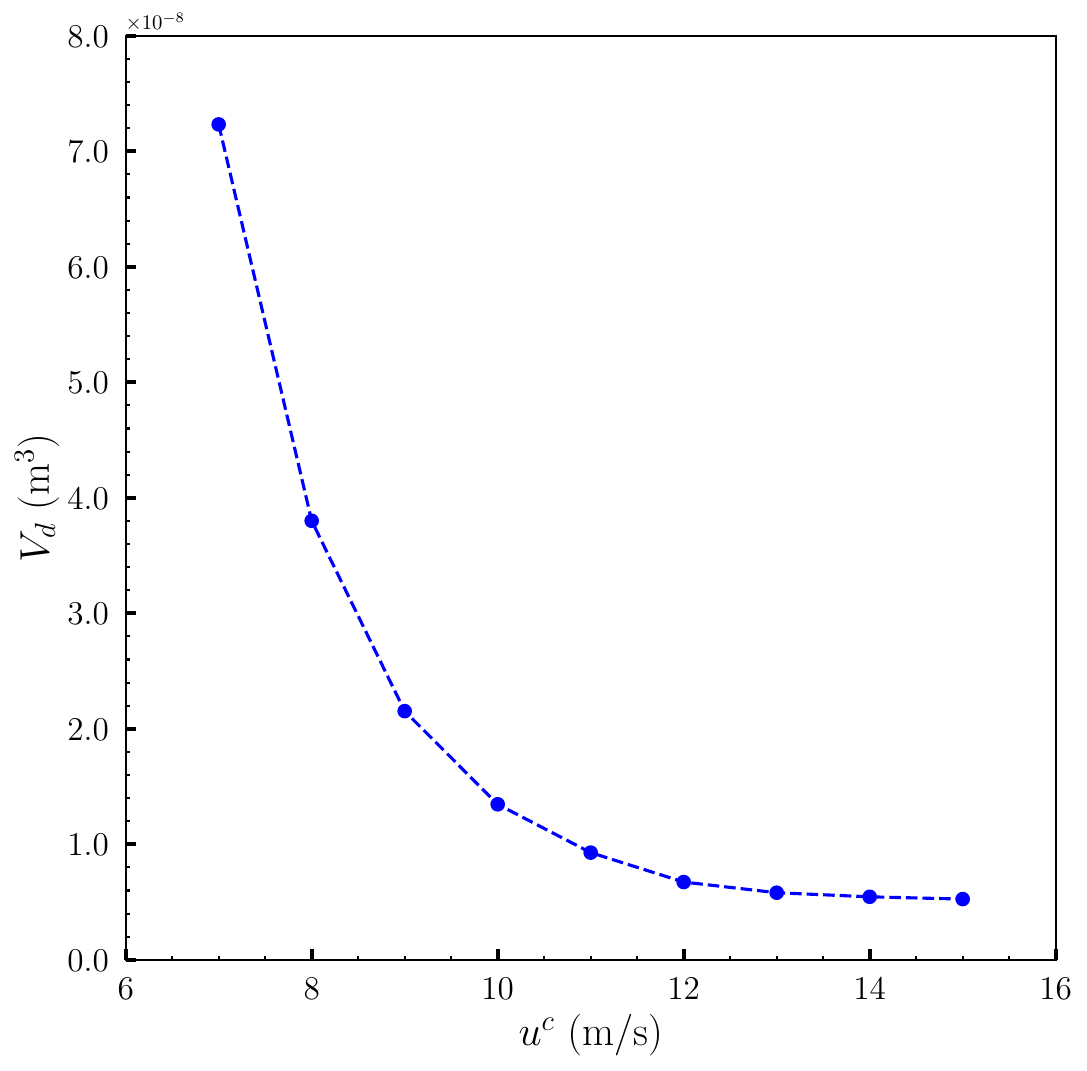}
      \caption{}
      \label{fig:wax_V_drop_shear}
    \end{subfigure}
    \caption{(a) Radius and (b) Volume of the primary droplet for various continuous phase velocities for shear-induced liquid paraffin wax droplet.}
\end{figure}

The chief quantities of interest for the shear-induced paraffin wax droplet formation are the pinch-off volume of the primary droplet and the pinch-off time. Fig.~(\ref{fig:wax_t_pinch_shear}) shows the pinch-off time for increasing continuous phase velocity. The pinch-off time decreases by almost a factor of $10$ from $u^c = 7$ m/s to $u^c = 15$ m/s. The continuous phase flow contributes to the forces in both normal and tangential directions on the interface. Increasing velocity increases normal and tangential forces that force the fluid column to approach the singularity while pushing it for more elongation. This reduces the overall time for radial expansion and necking, ultimately lowering the pinch-off time. This behavior is visible from the series of images in Figures (\ref{fig:uc_07} - \ref{fig:uc_15}). The radial expansion phase is clearly visible from $-0.5$ non-dimensional time away from pinch-off for continuous phase velocity $7$, $8$, and $9$,  which suggests that a lower velocity range for the continuous phase does not provide enough shear force to eliminate this inflation. This is why the pinch-off time is very high for these cases. For velocity $10$ m/s and higher, the shearing effect increases and creates a thinner fluid column. This helps the surface tension naturally in shrinking the fluid column radius, and it takes less time to get a pinch-off. Although the pinch-off time approaches a certain lower bound that suggests the minimum time for pinch-off for the dripping regime.

The droplet volume and droplet size are other quantities of interest for the analysis. Fig.~(\ref{fig:wax_h_drop_shear}) shows the primary droplet radius for increasing continuous phase velocity. The droplet radius decreases with the increasing shearing effect. Due to the high shearing effect, the thinner fluid column creates a smaller droplet. The images on the right in Figures (\ref{fig:uc_07} - \ref{fig:uc_15}) show the comparison of the primary droplet sizes. The volume of the pinch-off droplets also decreases because of the smaller droplet sizes. Fig.~(\ref{fig:wax_V_drop_shear}) shows the plot for decreasing pinch-off volume for increasing velocity. This is also due to less fluid accumulation in the column coming from the shearing effect. Since the higher outer velocity forces a faster pinch-off, the added volume from the inlet is smaller compared to that from a lower continuous phase velocity.
 
The pinch-off volume and pinch-off time are both decreasing with the increasing outer flow radius. However, the volume rate of pinching droplets per unit time increases because the pinch-off time decreases by a higher magnitude than the pinch-off volume. This is shown in Fig.~(\ref{fig:wax_V_t_shear}). For the lower range of the continuous phase velocity profile, the pinch-off volume rate increases slowly due to a higher pinch-off time. However, the increment in pinch-off volume rate is very fast after $u_c = 10$ m/s. The increment trend slows down for the higher droplet velocity range due to very little change in pinch-off time and volume.

\begin{figure}
    \centering
    \begin{subfigure}[t]{\linewidth}
      \centering
      \includegraphics[width=0.7\textwidth]{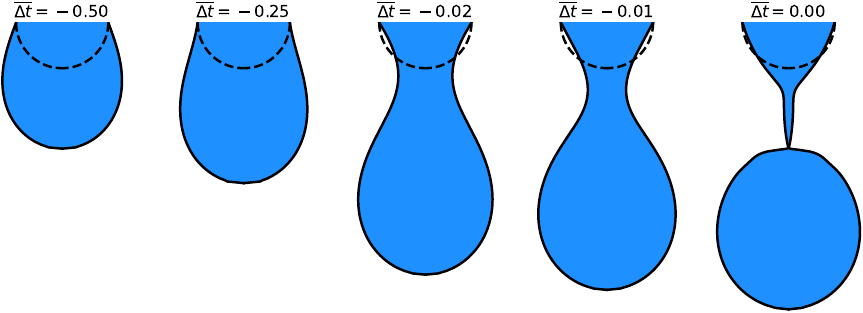}
      \caption{}
      \label{fig:uc_07}
    \end{subfigure}
    \begin{subfigure}[t]{\linewidth}
      \centering
      \includegraphics[width=0.7\textwidth]{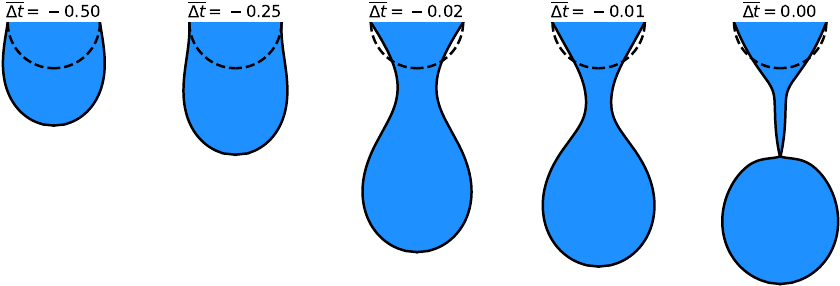}
      \caption{}
      \label{fig:uc_08}
    \end{subfigure}
    \begin{subfigure}[t]{\linewidth}
      \centering
      \includegraphics[width=0.7\textwidth]{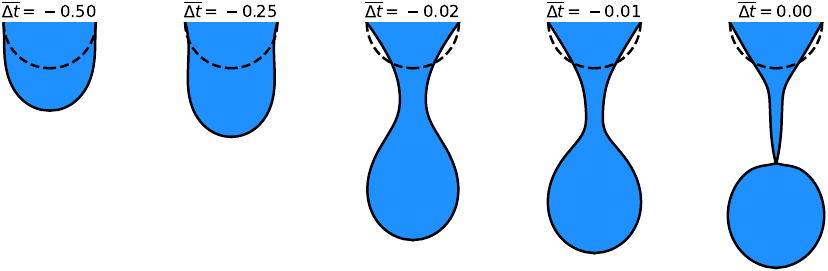}
      \caption{}
      \label{fig:uc_09}
    \end{subfigure}
    \begin{subfigure}[t]{\linewidth}
      \centering
      \includegraphics[width=0.7\textwidth]{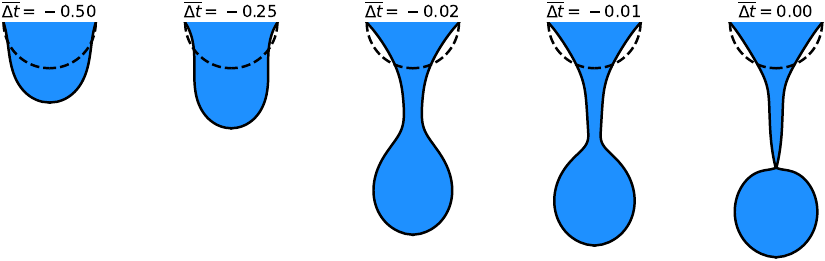}
      \caption{}
      \label{fig:uc_10}
    \end{subfigure}
    \begin{subfigure}[t]{\linewidth}
      \centering
      \includegraphics[width=0.7\textwidth]{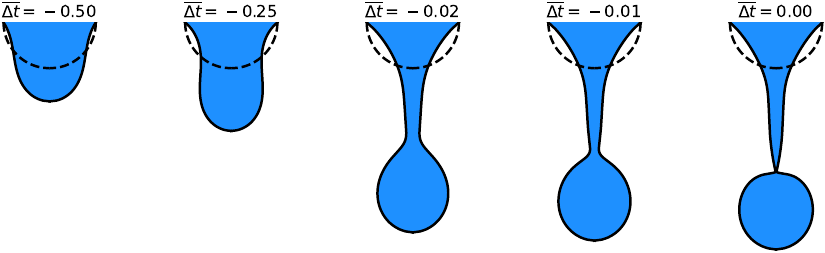}
      \caption{}
      \label{fig:uc_11}
    \end{subfigure}
\end{figure}
\begin{figure}
    \ContinuedFloat
    \centering
    \begin{subfigure}[t]{\linewidth}
      \centering
      \includegraphics[width=0.7\textwidth]{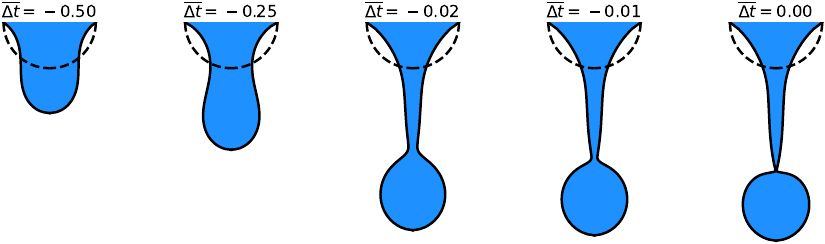}
      \caption{}
      \label{fig:uc_12}
    \end{subfigure}
    \begin{subfigure}[t]{\linewidth}
      \centering
      \includegraphics[width=0.7\textwidth]{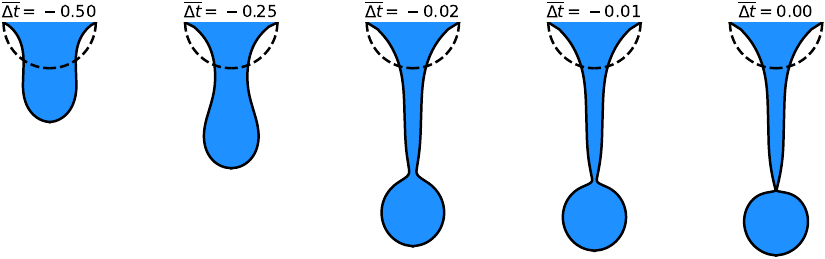}
      \caption{}
      \label{fig:uc_13}
    \end{subfigure}
    \begin{subfigure}[t]{\linewidth}
      \centering
      \includegraphics[width=0.7\textwidth]{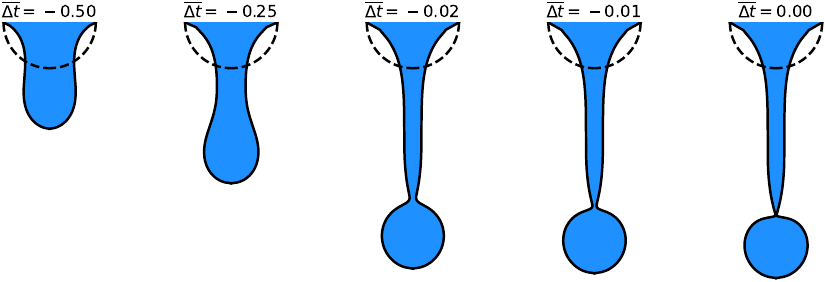}
      \caption{}
      \label{fig:uc_14}
    \end{subfigure}
    \begin{subfigure}[t]{\linewidth}
      \centering
      \includegraphics[width=0.7\textwidth]{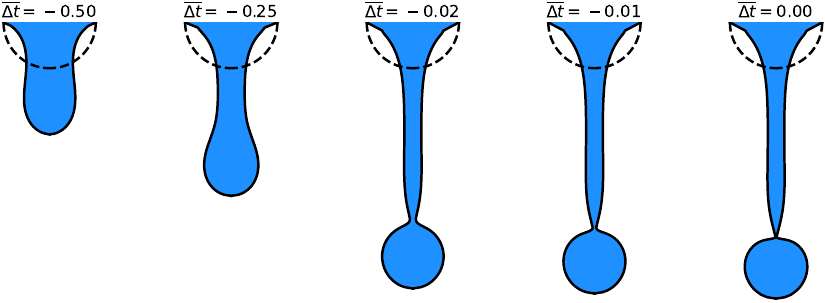}
      \caption{}
      \label{fig:uc_15}
    \end{subfigure}
    \caption{Droplet evolution until the primary pinch-off for the liquid paraffin wax in a shear force environment. The range of continuous phase flow velocity is $7$ m/s to $15$ m/s represented by (\ref{fig:uc_07}) to (\ref{fig:uc_15}). The profiles are snaps at a non-dimensional time away from the pinch-off.}
\end{figure}

\clearpage
\section{Conclusion}
The increasing importance of shear-induced droplet formation in multi-phase flow applications shows the need for efficient, high-fidelity modeling. We present a one-dimensional mathematical model that can be useful to simulate droplet formation in co-flowing fluids with a prescribed velocity for the continuous phase fluid. We propose that the continuous phase flow velocity, modeled as a Poiseuille flow, be used in the force balance on the interface, which simplifies the momentum equation. The interface is then advected with the flow. The discretization of the model equations is accomplished by a mixed finite element approach. We use PETSc~\cite{petsc-user-ref,petsc-web-page}, an open-source toolkit, to discretize and solve the equations. 

The proposed model consists of a single parameter that is defined using previous experimental and computational studies. This cross-validated definition of the parameter is then used to simulate liquid paraffin wax droplets in co-flowing airflow. The primary quantities of interest were droplet pinch-off volume and pinch-off time. Our simulation results reported decreasing pinch-off time for increasing continuous phase velocity. This happens due to the increased forces in both normal and tangential directions, which produce a more elongated structure at a faster rate. However, this decreasing trend in pinch-off time approaches a lower bound, suggesting a minimum time required for pinch-off, which depends on the viscous, inertial, and surface tension forces. We noticed decreasing droplet radius with increasing shear force from the outer fluid. The high shear force creates a thinner fluid column that produces smaller size droplets. This shows that the primary droplet volume reduction is due to the cubic relation with the radius. However, the pinch-off volume per unit time has an increasing trend for increasing shear force. 

The findings in this paper for paraffin wax droplets are particularly useful for predicting fuel utilization in the hybrid rocket combustion process. The regression of the paraffin-based fuel happens partially due to the atomization process, where droplet formation in a fast-moving oxidizer gas forms the ligaments of liquid paraffin wax, leading to droplet formation. We aim to apply this approach to predict the paraffin droplet sizes that can be used to estimate the regression rate due to atomization.

\section{Acknowledgement}
Funded by the United States Department of Energy’s (DoE) National Nuclear Security Administration (NNSA) under the Predictive Science Academic Alliance Program III (PSAAP III) at the University at Buffalo, under contract number DE-NA0003961.

This work was partially supported by the National Science Foundation under Grant No. NSF SI2-SSI: 1450339.
%
% \section{References}
% \bibliographystyle{elsarticle-harv} 
% \bibliography{ref}% Produces the bibliography via BibTeX.
\printbibliography
\end{document}